\documentclass{ws-procs961x669}            
\begin{document}
\title{Chiral magnetic effect in heavy ion collisions and beyond}

\author{Dmitri E. Kharzeev$^*$}

\address{Center for Nuclear Theory,\\ 
Department of Physics and Astronomy, Stony Brook University\\
New York 11794-3800, USA\\
Department of Physics, Brookhaven National Laboratory\\
Upton, New York 11973-5000, USA\\
$^*$E-mail: dmitri.kharzeev@stonybrook.edu\\
www.physics.sunysb.edu/kharzeev}

\begin{abstract}
Chirality is a ubiquitous concept in modern science, from particle physics to biology. In quantum physics, chirality of fermions is linked to topology of gauge fields by the chiral anomaly. While the chiral anomaly is usually associated with the short-distance behavior in field theory, in recent years it has been realized that it also affects the macroscopic behavior of systems with chiral fermions. In particular, the local imbalance between left- and right-handed fermions in the presence of a magnetic field induces non-dissipative transport of electric charge ("the Chiral Magnetic Effect", CME). In heavy ion collisions, there is an ongoing search for this effect at Relativistic Heavy Ion Collider, with results from a dedicated isobar run presented very recently. An observation of CME in heavy ion collisions could shed light on the mechanism of baryon asymmetry generation in the Early Universe. 
Recently, the CME has been discovered in Dirac and Weyl semimetals possessing chiral quasi-particles. This observation opens a path towards quantum sensors, and potentially a new kind of quantum computers.
\end{abstract}


\bodymatter

\section{Chirality in subatomic world}\label{aba:sec1}

In 1874, Louis Pasteur wrote \cite{pasteur}:
{\it ``The universe is asymmetric and I am persuaded that life, as it is known to us, is a direct result of the asymmetry of the universe or of its indirect consequences."} Life {\it is} asymmetric, and the concept of chirality -- the distinction between the left and right, or between an object and its reflection in the mirror -- is quite literally ingrained in our DNA. The DNA double helix is chiral; moreover,  DNA molecules also tend to form knots\cite{dna} -- and most of the complex knots are chiral as well. 

Pasteur's prophecy is that the chiral asymmetry of life originates from the asymmetry of the Universe. At present we still do not understand how this connection operates -- but we do  know already that the Universe is asymmetric. As predicted in 1956 by T.D. Lee and C.N. Yang \cite{Lee:1956qn}, and established shortly afterwards through the observation of angular asymmetry in $\beta$ decay by C.S. Wu \cite{Wu:1957my}, weak interactions involve only left-handed fermions. This discovery brought to focus the crucial role of chirality in subatomic world. 

For fermions, chirality is defined as the projection of spin on momentum, with a minus sign for antifermions. In Dirac theory, it is an eigenvalue of $\gamma^5$ matrix, with projectors on the right and left chiral states given by  
\begin{equation}\nonumber
{\cal P}^+ = \frac{1}{2} ({\bf 1} + \gamma^5)   ;\  \   \
{\cal P}^- = \frac{1}{2} ({\bf 1} - \gamma^5).
\end{equation} 
A Dirac spinor $\psi$ can thus be decomposed into right- and left-handed chiral states by $\psi^+ = {\cal P}^+ \psi$ and $\psi^- = {\cal P}^- \psi$. 
Dirac equation for massless fermions is $i{\hat D} \psi \equiv i \gamma^\mu \partial_\mu \psi = 0$, and Dirac operator can also  be decomposed into right- and left-handed components: 
\begin{equation}\label{dir}
D^+ \equiv i {\hat D} {\cal P}^+ ;\ \ \ D^- \equiv i {\hat D} {\cal P}^- . 
\end{equation}

For massless fermions, chirality is conserved in the interactions with gauge fields (e.g. photons and gluons) on the classical level - so a left-handed fermion is expected to always stay left-handed. However quantum chiral anomaly \cite{Adler:1969gk,Bell:1969ts} allows for a chirality transmutation -- a left-handed fermion can become right-handed if it interacts with a configuration of gauge field that can change its own chirality. While the chirality of a fermion  is a familiar concept, how does one characterize chirality of a gauge field?

\section{Chirality of gauge fields}

It appears that the key to understanding the chirality of gauge fields is offered by the example of DNA knots mentioned above. Indeed, let us consider a knot of magnetic flux. The chiral invariant of this gauge field configuration is known as {\it magnetic helicity}: 
\begin{equation}\label{cs}
{\rm CS}[A] = \int d^3x\ {\bf A} \cdot {\bf B},
\end{equation}  
where ${\bf A}$ is the vector gauge potential, and ${\bf B} = {\bf \nabla} \times {\bf A}$ is magnetic field. 
Using analogy between the gauge potential and velocity $\bf v$, we can easily understand that (\ref{cs}) detects chirality -- indeed, when ${\bf A} \to {\bf v}$, magnetic field gets replaced by vorticity ${\bf B} \to {\bf \Omega} = \frac{1}{2} {\bf \nabla} \times {\bf v}$,
and a non-zero scalar product ${\bf v} \cdot {\bf \Omega}$ implies a helical motion.

For an Abelian gauge theory (such as electrodynamics), magnetic helicity coincides with a more general 3-form derived by Chern and Simons \cite{Chern:1974ft} in differential geometry to characterize the global topology of a manifold with a Lie algebra valued 1-form ${\bf A}$ over it:
\begin{equation}\label{csw}
{\rm CS}[A] = {\rm Tr} \left[ {\bf F} \wedge {\bf A} - \frac{1}{3} {\bf A} \wedge {\bf A} \wedge {\bf A} \right],
\end{equation} 
where the curvature (field strength tensor) is defined as ${\bf F} = d {\bf A} + {\bf A} \wedge {\bf A}$. 
In physical terms, the invariant 
(\ref{cs}) corresponds to the chirality of the knot made of magnetic flux - for example, a torus would be characterized by ${\rm CS} = 0$, a right-handed trefoil knot by ${\rm CS} = +1$, and a left-handed one by ${\rm CS} = -1$. 

Chern-Simons $p$-form can be defined for any odd space-time dimension $p$. We happen to live in $3+1$ dimensional space-time with even $p=4$ -- does this mean that Chern-Simons invariant is irrelevant? Of course not -- what this means is that in our four-dimensional space-time Chern-Simons invariant can change in time, with non-vanishing exterior derivative 
\begin{equation}\label{csw}
d {\rm CS}[A] = {\rm Tr} \left[ {\bf F} \wedge {\bf F} \right],
\end{equation} 
which is known as the Chern-Pontryagin invariant. 
In a more familiar for physicists notation, this relation implies that 
\begin{equation}\label{cst}
\frac{\partial\ {\rm CS}[A]}{\partial t} = - 2 \int d^3x\ {\bf E} \cdot {\bf B}, 
\end{equation} 
which can be readily obtained from (\ref{cs}) using the Coulomb gauge where the electric field ${\bf E}$ is related to the gauge potential by ${\bf E} = - {\dot{\bf A}}$. Therefore, a change of topology of the gauge field in time quantified by (\ref{cst}) gives rise to an electric field parallel to the magnetic one. What is the effect of parallel electric and magnetic fields on chirality of chiral fermions? Can this electric field drive a current?

\section{Chiral anomaly}

Charged particles experience a Lorentz force ${\bf F}_m = e\ {\bf v} \times {\bf B}$ from an external magnetic field ${\bf B}$. If the projection of their velocities on the direction of magnetic field is equal to zero, this leads to the motion along closed cyclotron orbits with ${\bf \Omega} = {\bf \nabla} \times {\bf v} \neq 0$, but ${\bf v} \cdot {\bf \Omega} = 0$. Even if the charge has a non-zero velocity component along ${\bf \Omega}$, but there is no external force directed along ${\bf B}$, then we can always choose a frame in which ${\bf v} \cdot {\bf B} = {\bf v} \cdot {\bf \Omega} = 0$, i.e. the motion of the charge is not helical. This is not true if there is a force applied along the direction of ${\bf B}$, e.g. a Lorentz force ${\bf F}_e = e\ {\bf E}$ resulting from an electric field ${\bf E}$ parallel to ${\bf B}$ - then the motion is helical in any inertial frame.

In quantum theory, charged particles occupy quantized Landau levels, and for massless fermions the lowest Landau level (LLL) is chiral and has a zero energy -- qualitatively, this appears due to a cancellation between a positive kinetic energy and negative Zeeman energy of the interaction between magnetic field and spin. Therefore, on the LLL the spins of positive (negative) fermions are aligned along (against) the direction of magnetic field. All excited levels are degenerate in spin, and are thus not chiral. 

More formally, this can be seen as a consequence of Atiyah-Singer index theorem \cite{Atiyah:1968mp} that relates the analytical index of Dirac operator to its topological index. In other words, it relates the number of zero modes of Dirac operator acting on a manifold $M$ to the topology of this manifold. The analytical index of Dirac operator (\ref{dir}) is given by the number of zero energy modes with right ($\nu_+$) and left ($\nu_+$) chirality:
\begin{equation}
{\rm ind}\ D = {\rm dim\ ker} D^+ - {\rm dim\ ker} D^- = \nu_+ - \nu_- ,
\end{equation}
where $\ker D$ is the subspace spanned by the kernel of the operator $D$, i.e. the subspace of states that obey $D^+ \psi = 0$, or $D^- \psi = 0$ see (\ref{dir}). 

For two dimensional manifold $M$, the topological index of this operator is equal to $\frac{1}{2\pi} \int_M {\rm tr}\ {\bf F}$, and Atiyah-Singer index theorem thus states that
\begin{equation}\label{asi}
\nu_+ - \nu_- = \frac{1}{2\pi} \int_M {\rm Tr}\ {\bf F} .
\end{equation}
Performing analytical continuation to Euclidean ($x,y$) space (with ${\bf B}$ along the ${\bf z}$ axis), we thus find that the number of chiral zero modes is given by the total magnetic flux through the system \cite{Aharonov:1978gb}. For positive fermions with charge $e>0$, we have $\nu_-=0$ and the number of right-handed chiral modes from (\ref{asi}) is given by
\begin{equation}
\nu_+ = \frac{e \Phi}{2 \pi} ,
\end{equation}
which is just the number of LLLs in the transverse plane; we have included an explicit dependence on electric charge $e$. For negative fermions $\nu_+=0$ and $\nu_- = \frac{e \Phi}{2 \pi}$.

Let us assume for simplicity that the charge chemical potential is equal to zero, $\mu = 0$. It is clear that $\nu_+ = \nu_-$, and the system possesses zero chirality. Let us now turn on an external electric field ${\bf E} \parallel {\bf B}$. Dynamics of fermions on the LLL is $(1+1)$ dimensional along the direction ${\bf B}$, and we can apply the index theorem (\ref{asi}) to the $(z,t)$ manifold; for positive fermions of $"+"$ chirality
\begin{equation}\label{asi-ep}
\nu_+ = \frac{1}{2\pi} \int dz dt\ e E, \ \ \ \nu_- = 0,
\end{equation}
and for negative fermions of $"-"$ chirality
\begin{equation}\label{asi-en}
\nu_+ = 0, \ \ \ \nu_- = - \frac{1}{2\pi} \int dz dt\ e E .
\end{equation}
These relations can be understood from a {\it seemingly} classical argument \cite{Nielsen:1983rb}: the positive charges are accelerated by the Lorentz force along the electric field ${\bf E}$, and acquire Fermi-momentum $p_F^+ = e E t$. The density of states in one spatial dimension is $p_F/(2 \pi)$, and so the total number of positive fermions with positive chirality is $\nu_+ = 1/(2\pi) \int dz dt\ e E$, in accord with (\ref{asi-ep}). The same argument applied to negative fermions explains (\ref{asi-en}). While the notion of acceleration by Lorentz force is classical, in assuming that it increases the Fermi momentum, we have made an 
assumption that there is an infinite tower of states that are accelerated by the Lorentz force. This tower of states does not exist in classical theory; however it is a crucial ingredient of the (quantum) theory of Dirac.

Multiplying the density of states in longitudinal direction $p_F/(2 \pi)$ by the density of states $eB/(2 \pi)$ in the transverse direction, we find from (\ref{asi-ep}) and (\ref{asi-en}) that in $(3+1)$ dimensions
\begin{equation}\label{asi3}
\nu_+ - \nu_- = 2 \times \frac{e^2}{4\pi^2} \int d^2x\ dz\ dt\ {\bf E\cdot B}  = \frac{e^2}{2\pi^2} \int d^2x\ dz\ dt\ {\bf E\cdot B},
\end{equation}
where the factor of 2 is due to the contributions of positive and negative fermions. This relation represents Atiyah-Singer theorem for $U(1)$ in $(3+1)$ dimensions, so we could use it directly instead of relying on dimensional reduction of the LLL dynamics. Note that the quantity on the r.h.s. of (\ref{asi3}) is nothing but the derivative of Chern-Simons three-form, see (\ref{csw}), (\ref{cst}). 

The relation (\ref{asi3}) can also be written in differential form in terms of the axial current
\begin{equation}
J_\mu^5 = {\bar \psi} \gamma_\mu \gamma^5 \psi = J_\mu^+ - J_\mu^-
\end{equation}
as \cite{Adler:1969gk,Bell:1969ts} 
\begin{equation}\label{an-d}
\partial^\mu J_\mu^5 =  \frac{e^2}{2\pi^2}\ {\bf E\cdot B}.
\end{equation}
The chiral anomaly equation (\ref{an-d}) is an operator relation. In particular, we can use it to evaluate the matrix element of transition from a pseudoscalar excitation of Dirac vacuum (a neutral pion) into two photons. This can be done by using on the l.h.s. of (\ref{an-d}) the PCAC relation that replaces the divergence of axial current by the interpolating pion field $\varphi$
\begin{equation}
\partial^\mu J_\mu^5 \simeq F_\pi M_\pi^2\ \varphi,
\end{equation}
where $F_\pi$ and $M_\pi$ are the pion decay constant and mass. 
Taking the matrix element of (\ref{an-d}) between the vacuum and the two-photon states $\langle 0| \partial^\mu J_\mu^5 |\gamma \gamma \rangle$ then yields the decay width of $\pi^0 \to \gamma  \gamma$ decay which is a hallmark of chiral anomaly. However, the chiral anomaly has much broader  implications when the classical gauge fields are involved, as we will now discuss.

\section{Chiral magnetic effect:\\ non-dissipative quantum transport induced by chiral anomaly}

As a first step, let us observe that the derivation of chiral anomaly presented above implies the existence of non-dissipative electric current in parallel electric and magnetic fields. Indeed, the (vector) electric current 
\begin{equation}
J_\mu = {\bar \psi} \gamma_\mu  \psi = J_\mu^+ + J_\mu^-
\end{equation}
contains equal contributions from positive charge, positive chirality fermions flowing along the direction of $\bf E$ (which we assume to be parallel to $\bf B$), and negative charge, negative chirality fermions flowing in the direction opposite to $\bf E$:
\begin{equation}\label{cur-a}
J_z = 2 \times \frac{e^2}{4\pi^2} {\bf E\cdot B}\ t  = \frac{e^2}{2\pi^2} {\bf E\cdot B}\ t .
\end{equation}
In constant electric and magnetic fields, this current grows linearly in time -- this means that the conductivity $\sigma$ defined by $J = \sigma E$ becomes divergent, and resistivity $\rho = 1/\sigma$ vanishes. This means that the current (\ref{cur-a}) is non-dissipative, similarly to what happens in superconductors!

We can also write down the relation (\ref{cur-a}) in terms of the chemical potentials $\mu_+ = p_F^+$ and $\mu_- = p_F^-$ for right- and left-handed fermions, which for massless dispersion relation are given by the corresponding Fermi momenta $p_F^+ = e E t$ and $p_F^- = - e E t$. It is convenient to define the {\it chiral chemical potential} 
\begin{equation}
\mu_5 \equiv \frac{1}{2} \left( \mu_+ - \mu_- \right) 
\end{equation}
related to the density of chiral charge $\rho_5 = J_0^5$; for small $\mu_5$, it is proportional to $\rho_5$, $\mu_5 \simeq \chi^{-1} \rho_5$ where $\chi$ is the chiral susceptibility.
The relation (\ref{cur-a}) then becomes \cite{Fukushima:2008xe}
\begin{equation}\label{cme}
{\bf J} = \frac{e^2}{2\pi^2}\ \mu_5\ {\bf B} .
\end{equation}
It is important to realize that unlike a usual chemical potential, the chiral chemical potential $\mu_5$ does not correspond to a conserved quantity -- on the contrary, the non-conservation of chiral charge due to chiral anomaly is necessary for the  Chiral Magnetic Effect (CME) described by (\ref{cme}) to exist. Indeed, a static magnetic field cannot perform work, so the current (\ref{cme}) can be powered only by a change in the chiral chemical potential. Another way to see this is to consider a power \cite{Basar:2013iaa} of  the current (\ref{cme}): $P = \int d^3x\ {\bf E}\ {\bf J} \sim \mu_5\ \int d^3x\ {\bf E}\ {\bf B}$. For a constant $\mu_5$, it can be both positive or negative, in clear contradiction with energy conservation. In particular, one would be able to extract energy from the ground state of the system characterized by $\mu_5 \neq 0$! On the other hand, if $\mu_5$ is dynamically generated through the chiral anomaly (\ref{an-d}), it has the same sign as $\int d^3x\ {\bf E}\ {\bf B}$, and the electric power is always positive, as it should be. It is important to note that in the latter case the state with $\mu_5 \neq 0$ is not the ground state of the system, and can relax to the true ground state through the anomaly by generating the CME current.

For the case of parallel ${\bf E}$ and ${\bf B}$, the CME relation (\ref{cme}) is a direct consequence of the Abelian chiral anomaly for the case of classical background fields. However it is valid also when the chiral chemical potential is sourced by non-Abelian anomalies \cite{Fukushima:2008xe}, coupling to time-dependent axion field \cite{Wilczek:1987mv}, or is just a consequence of a non-equilibrium dynamics \cite{Kharzeev:2016mvi}.

The relation (\ref{cme}) was first introduced in 1980 in a pioneering paper by Vilenkin \cite{Vilenkin:1980fu} who however considered the case of a constant $\mu_5$, motivated by parity violation in weak interactions. In this case, as we have discussed above, the system is in equilibrium and the electric current cannot exist \cite{Vilenkin:1980ft} -- so while the relation (\ref{cme}) was known for some time, turning it into a real physical effect required understanding the role of anomaly, see review \cite{Kharzeev:2013ffa} for a detailed discussion and references. The same is true in condensed matter applications when the nodes of dispersion relations of left- and right-handed fermions are located at different energies, $E_+(p=0) \neq E_-(p=0)$. In this case, contrary to some early claims, the CME current does not appear \cite{Basar:2013iaa}. For the CME to emerge, the chiral asymmetry has to appear in the occupancy of the left- and right-handed states.

\section{Chiral magnetic effect as a probe of vacuum topology:\\ interplay of Abelian and non-Abelian chiral anomalies}

The CME relation (\ref{cme}) can be proven also for the case of non-Abelian plasma containing chiral fermions in an external Abelian magnetic field ${\bf B}$ \cite{Fukushima:2008xe}. In this case, the chiral charge (and the corresponding chiral chemical potential) is created by non-Abelian anomaly due to transitions \cite{Belavin:1975fg} between different topological sectors marked by different Chern-Simons numbers (\ref{csw}), and the electric current flows along (or against) the direction of ${\bf B}$. This presents a unique opportunity to get a direct experimental access to the study of non-Abelian topological fluctuations \cite{Kharzeev:2004ey,Kharzeev:2007jp,Kharzeev:2007tn}. Topological transitions in non-Abelian electroweak plasma violate the baryon number conservation and may be at the origin of baryon asymmetry of our Universe \cite{Kuzmin:1985mm,Rubakov:1996vz}. 

The only non-Abelian theory where topological transitions are accessible to experiment is Quantum ChromoDynamics (QCD). Its chiral fermions are quarks, that are confined into hadrons. According to the Atiyah-Singer index theorem \cite{Atiyah:1968mp}, topological transitions in QCD should be accompanied by the change in chirality of quarks. They can thus be responsible for spontaneous breaking of ${\rm U_+(3) \times U_-(3)}$ chiral symmetry in QCD, emergence of the corresponding Goldstone bosons (pions, kaons, and $\eta$ meson), and a massive flavor-singlet $\eta'$ meson resulting from an explicit breaking of the flavor-singlet part of the chiral symmetry. 

While there is a significant evidence for the prominent role of topological transitions (instantons, sphalerons, ...) in the structure of QCD vacuum and the properties of hadrons (see \cite{Schafer:1996wv} for a review), such transitions have never been detected in experiment. The CME, with its directly detectable electric current,  makes such an observation possible. Indeed, consider a QCD plasma is created in a heavy ion collision. Because the colliding ions possess positive electric charges, the produced plasma, at least during its early moments, is embedded into a very strong magnetic field
which is on the order of a typical QCD scale \cite{Kharzeev:2007jp}, $eB \sim \Lambda_{\rm QCD}^2$. In such a strong magnetic field (possibly the strongest in the present Universe), electromagnetic interactions of quarks are comparable in strength to the strong ones. In addition, the rate of topological ``sphaleron" \cite{Klinkhamer:1984di} transitions in hot QCD plasma is high, and this should result in the creation of chirally imbalanced domains (``${\cal P}$-odd bubbles" \cite{Kharzeev:1998kz}) characterized by non-zero value of the chiral chemical potential. This means that all conditions for the CME are met, and there should be an electric current (\ref{cme}) propagating through the QCD plasma.

\section{Chiral magnetic effect in heavy ion collisions\\ and the RHIC isobar run}

The magnetic field produced by the colliding ions is directed perpendicular to the reaction plane of the collision, therefore the CME current (\ref{cme}) is directed perpendicular to the reaction plane as well. The magnetic field is strongest during the early moments of the collision; the sphaleron transitions are thus expected to induce the electric charge separation relative to the reaction plane early in the evolution of the system \cite{Kharzeev:2004ey}. Because the QCD plasma is rapidly expanding (similarly to the Early Universe), this initial electric charge separation cannot be fully scrambled by final state interactions, and survives till the moment when the plasma cools down and transforms into hadrons. Since the electric charge is conserved throughout the hadronization, the produced hadrons should possess charge asymmetry relative to the reaction plane that can be directly detected in experiment. 

Of course, QCD does not violate parity ${\cal P}$ symmetry globally, and the sign of chiral imbalance fluctuates event-by-event -- we are thus dealing with a ``local ${\cal P}$-violation". The experimental signature of CME is thus a dynamical enhancement of out-of-plane fluctuations of charge asymmetry, relative to the in-plane fluctuations that are not affected by CME. The corresponding experimental observable was proposed by Voloshin \cite{Voloshin:2004vk} almost immediately after the idea to search for CME in heavy ion collisions was formulated \cite{Kharzeev:2004ey}. In terms of the azimuthal angles of positive $\phi^+$ and negative $\phi^-$ hadrons, and the azimuthal angle of the reaction plane $\Psi_{RP}$ the proposed ``$\gamma$ correlator" observable is \cite{Voloshin:2004vk}
\begin{equation}\nonumber
\gamma^{+-} \equiv \left\langle \cos( \phi^+ + \phi^- - 2 \Psi_{RP}) \right\rangle = 
\end{equation}
\begin{equation}\label{obs}
= \left\langle \cos( \phi^+ - \Psi_{RP}) \cos( \phi^- - \Psi_{RP}) \right\rangle - \left\langle \sin( \phi^+ - \Psi_{RP}) \sin( \phi^- - \Psi_{RP}) \right\rangle,
\end{equation}
where the sum over hadrons in a given event, and then an average over many events, are assumed. 

The CME would result in the electric charge separation relative to the reaction plane, and should produce, in a given event, either $\sin( \phi^+ - \Psi_{RP}) > 0$, $\sin( \phi^+ - \Psi_{RP}) < 0$, or $\sin( \phi^+ - \Psi_{RP}) < 0$, $\sin( \phi^+ - \Psi_{RP}) > 0$, depending on the sign of the chiral imbalance. In both cases however $\left\langle \sin( \phi^+ - \Psi_{RP}) \sin( \phi^- - \Psi_{RP}) \right\rangle < 0$, and the correlator (\ref{obs}) should be positive. Of course, background fluctuations would also give contributions to the $\gamma$ correlator, but (\ref{obs}) is the difference of in-plane and out-of-plane fluctuations, so the backgrounds that do not depend on the reaction plane cancel out. Therefore the backgrounds that survive in (\ref{obs}) should depend on the reaction plane, and are thus expected to be proportional to the ``elliptic flow" that describes the elliptic deformation of the event and is defined as the second Fourier harmonic of the hadron azimuthal angle distribution \cite{Voloshin:2004vk}. In hydrodynamical description of heavy ion collisions, the elliptic flow results from the pressure anisotropy generated by the geometry of an off-central collision. It is also important to note that (\ref{obs}) is expected to scale with inverse hadron multiplicity, both for the signal and background contributions \cite{Kharzeev:2004ey,Voloshin:2004vk}, see \cite{Kharzeev:2015znc} for a review and detailed discussion.

STAR Collaboration performed the measurements of $\gamma$ correlators, and other CME observables, over a range of collision energies and for different ions \cite{STAR:2009wot,STAR:2009tro,STAR:2013ksd}. In addition, ALICE \cite{ALICE:2012nhw} and CMS \cite{CMS:2017lrw} Collaborations at the LHC extended these studies to higher energies, see \cite{Kharzeev:2015znc,Kharzeev:2020jxw} for reviews and compilations of published results. For all studied heavy ion colllisions, non-zero and positive $\gamma^{+-}$ was measured. The correlators $\gamma^{++}$ and $\gamma^{--}$ were also non-zero and negative, as expected for CME. However,  background contributions proportional to elliptic flow have also also identified, and the problem of separating the possible signal from background has assumed the center of the stage.

This is why the proposal of using isobar collisions has been made \cite{Voloshin:2010ut}. The idea is that since the isobars have the same mass number, they have {\it approximately} the same size and shape, and thus the elliptic flow, and the backgrounds driven by it, should be nearly identical. On the other hand, the difference in the electric charge should create a difference in the produced magnetic field, and thus in the CME which is proportional to it. The measurement of $\gamma$ correlators, and other CME observables, in isobar collisions would thus allow to isolate the CME signal from the background. 

A dedicated high statistics measurement of CME observables in ${\rm ^{96}_{44}Ru\ ^{96}_{44}Ru}$ and ${\rm ^{96}_{40}Zr\ ^{96}_{40}Zr}$ collisions was performed by STAR Collaboration at RHIC in 2018 \cite{STAR:2021mii}. The data from 3.8 billion collision events were recorded, and a blind analysis of the data (unprecedented in heavy ion physics) was performed. The expectation has been that since the electric charge of ${\rm Ru}$ is higher than that of ${\rm Zr}$, the ratio of CME observables, e.g. $\gamma^{Ru}/\gamma^{Zr}$ would exceed one if CME were present, and be equal to one if CME were to give a negligible contribution. In any scenario, one expects to find $\gamma^{Ru}/\gamma^{Zr} \geq 1$. 

The STAR analysis however revealed that the ratio $\gamma^{Ru}/\gamma^{Zr}$ was significantly below one! This certainly did not fit the CME expectations -- and in fact {\it any} theory expectation, CME-based or not. Basing on these findings, and the predefined criteria assuming the identical backgrounds in ${\rm ^{96}_{44}Ru\ ^{96}_{44}Ru}$ and ${\rm ^{96}_{40}Zr\ ^{96}_{40}Zr}$ collisions, STAR Collaboration concluded that ``no CME signature that satisfies the predefined criteria has been observed" \cite{STAR:2021mii}. 

While the original CME expectations for the isobar run have certainly not been met, the puzzling observation of $\gamma^{Ru}/\gamma^{Zr} < 1$ begs for an explanation. A detailed investigation is still ongoing, but the post-blinding analysis \cite{STAR:2021mii} has already revealed that the key to the solution is a very significant difference in hadron multiplicity  in ${\rm ^{96}_{44}Ru\ ^{96}_{44}Ru}$ and ${\rm ^{96}_{40}Zr\ ^{96}_{40}Zr}$ collisions observed by STAR in centrality cuts relevant for the CME analysis. As mentioned above, the $\gamma$ correlator scales, in the good first approximation, with inverse multiplicity, and the multiplicity measured in  ${\rm ^{96}_{44}Ru\ ^{96}_{44}Ru}$ is significantly higher than in ${\rm ^{96}_{40}Zr\ ^{96}_{40}Zr}$ collisions in the same centrality cut.
This explains the surprisingly low ratio $\gamma^{Ru}/\gamma^{Zr} < 1$. If one establishes a new baseline given by the measured ratio of inverse multiplicities, the observed $\gamma^{Ru}/\gamma^{Zr}$ ratio  in fact exceeds the baseline by $(1-4)\ \sigma$, depending on the details of the analysis. This supports the CME interpretation, albeit with an insufficient statistical significance. 

More detailed analysis of the isobar data taking account of the difference in the shape of Ru and Zr nuclei is still ongoing, but it has already become clear that the isobar run data represent an important milestone in the hunt for CME in heavy ion collisions. Nevertheless, the final conclusions will have to be based on the joint analysis of both isobar and symmetric heavy ion collisions. It will also be very important to extend the CME studies to lower collision energies, where topological fluctuations can be strongly enhanced \cite{Ikeda:2020agk} due to proximity to the critical point in the QCD phase diagram. This program is planned for the beam energy scan at RHIC, and can also be performed at future heavy ion facilities, such as NICA and FAIR.

\section{Broader implications}

CME is a macroscopic quantum phenomenon driven by the chiral anomaly; it is a direct probe of gauge field topology. As a result, it is induces a variety of phenomena involving chiral fermions in quark-gluon plasma, the Early Universe \cite{Brandenburg:2017rcb}, astrophysics \cite{Gorbar:2021tnw},  and condensed matter physics. In the latter case, the CME has been firmly established through the studies of magnetotransport in Dirac and Weyl semimetals  \cite{Li:2014bha,Ong}, and is finding new applications. The emerging new frontier in condensed matter physics is chiral photonics, including the studies of chiral magnetic photocurrents \cite{Kaushik:2018tjj} and other chiral phenomena. Among potential future directions, let us mention the use of CME for controlling ``chiral qubits" \cite{Kharzeev:2019ceh} in quantum processors. It is clear that the chiral asymmetry of the Universe has many far-reaching consequences that we are just beginning to uncover.
\vskip0.3cm

This work was supported by the U.S. Department of Energy,
Office of Science grants No.
DE-FG88ER40388 and DE-SC0012704, and Office of Science, National Quantum
Information Science Research Centers, Co-design Center for Quantum Advantage (${\rm C^2QA}$) under contract number DE-SC0012704.

\end{document}